\documentstyle[epsf,12pt]{article}
\newcommand{\ol}{\overline}
\newcommand{\wt}{\widetilde}
\def\diag{\mathop{\rm diag}}
\def\sign{\mathop{\rm sign}}
\def\tr{\mathop{\rm tr}}
\def\mod{\mathop{\rm mod}}

\newcommand{\AD}[1]{$\ol{\mbox{D}{#1}}$}

\begin{document}

%%%%%%%%%%%%%%%%%%%%%%%%%%%%%%%%%%%%%%%%%%%%%%%%%%%%%%%%%%

				%titlepage

%%%%%%%%%%%%%%%%%%%%%%%%%%%%%%%%%%%%%%%%%%%%%%%%%%%%%%%%%%
\begin{titlepage}

    \begin{normalsize}
     \begin{flushright}
                 YITP-00-35\\
                 MIT-CTP-2994\\
                 hep-th/0007012\\
                 July 2000
     \end{flushright}
    \end{normalsize}
    \begin{Large}
       \vspace{0.6cm}
       \begin{center}
       \bf
        Orientifold Planes, Type I Wilson Lines\\
        and Non-BPS D-branes 
       \end{center}
    \end{Large}
  \vspace{1mm}

\begin{center}
           Yoshifumi Hyakutake
           \footnote{E-mail :
              hyaku@yukawa.kyoto-u.ac.jp},
           Yosuke Imamura
           \footnote{E-mail :
              imamura@hep-th.phys.s.u-tokyo.ac.jp\\
              \quad\quad Address from July 2000 :\\
              \quad\quad Detartment of Physics,
              University of Tokyo,
              Hongo, Tokyo 113-0033, Japan}
           and $\;$ Shigeki Sugimoto 
           \footnote{E-mail :
              sugimoto@yukawa.kyoto-u.ac.jp}  \\
      \vspace{6mm}
        ${}^{1,3}${\it Yukawa Institute for Theoretical Physics} \\
              {\it Kyoto University}\\
              {\it Sakyo-ku, Kyoto 606-8502, Japan}\\
       \vspace{4mm}
        ${}^{2}${\it Center for Theoretical Physics,} \\
                {\it Massachusetts Institute of Technology}\\
                {\it Cambridge MA 02139, USA}\\             
      \vspace{1cm}

    \large{ABSTRACT}
        \par
\end{center} 
\begin{quote}
\begin{normalsize}
%\begin{abstract}
There is a longstanding puzzle concerned with
the existence of $\wt{\mbox{O}p}^-$-planes with $p\ge 6$, which are
orientifold
$p$-planes of negative charge with stuck D$p$-branes.
We study the consistency of configurations with various
orientifold planes and propose a resolution to this puzzle.
It is argued
that $\wt{\mbox{O}6}^-$-planes are possible in massive IIA theory
with odd cosmological constant,
while $\wt{\mbox{O}7}^-$-planes and $\wt{\mbox{O}8}^-$-planes are not
allowed.
Various relations between orientifold planes and non-BPS D-branes
are also addressed.
\end{normalsize}
\end{quote}
%\end{abstract}

\end{titlepage}

%%%%%%%%%%%%%%%%%%%%%%%%%%%%%%%%%%%%%%%%%%%%%%%%%%%%%%%%%%%%%%%%%%%%%%%%%%
\section{Introduction}\label{intro.sec}

In type II string theory there are perturbative ${\bf Z}_2$
symmetries 
with orientation reversal on the world-sheet of fundamental 
strings. And by keeping only states invariant under these
${\bf Z}_2$ 
symmetries, we can construct consistent theories with fixed
planes called orientifold planes.
An orientifold plane extended in $p$ spatial directions
(and one timelike direction) is denoted by O$p$-plane.

The manifold enclosing an O$p$-plane is ${\bf RP}^{8-p}$ and
O$p$-planes are classified
by the cohomology groups of this manifold 
\cite{WittenBaryon,Hori,HananyKol}.
As an example, let us consider an O$5$-plane.
An O$5$-plane is surrounded by 
 ${\bf RP}^3$, which has the following non-trivial
integral cohomology groups
\begin{equation}
H^3({\bf RP}^3,{\bf Z})={\bf Z},\quad
H^3({\bf RP}^3,\wt{\bf Z})={\bf Z}_2,\quad
H^1({\bf RP}^3,\wt{\bf Z})={\bf Z}_2,
\end{equation}
where $\wt {\bf Z}$ is a ``twisted'' sheaf of integers.
The first one is associated with the R-R charge carried by the
O$5$-plane, i.e. the number of D5-branes coinciding it.
The second one is the discrete torsion associated with the NS
$B$-field and
this determines the sign of R-R charge of the O$5$-plane itself.
The unit element and the non-trivial element of
$H^3({\bf RP}^3,\wt{\bf Z})$ correspond to
the O$5$-plane with the negative and the positive R-R charge, respectively.
We will represent this charge by superscript like O$5^\pm$.
The third one is the discrete torsion associated with the R-R $0$-form
field (the axion field).
We use a notation $\wt{\mbox{O}5}$ for O$5$-planes associated with
a non-trivial element of this ${\bf Z}_2$ discrete torsion.

Similarly, for O$p$-planes with $p\le 5$,
there are two ${\bf Z}_2$ discrete torsions
associated
with the NS $B$-field and the R-R $(5-p)$-form field.
Therefore, there are (at least) four kinds of orientifold $p$-planes,
O$p^\pm$ and $\wt{\mbox{O}p}^\pm$.
The R-R charges 
and the gauge groups which appear on these orientifold
planes with $n$
D$p$-branes are as follows.
$$
\footnotesize
\begin{array}{|c|cccc|}
\hline
 & \mbox{O}p^- & \mbox{O}p^+ & \wt{\mbox{O}p}^- & \wt{\mbox{O}p}^+\\
\hline
\mbox{R-R charge}&n-2^{p-5}&n+2^{p-5}&n+1/2-2^{p-5}&n+2^{p-5}\\
\mbox{gauge group}& SO(2n) & USp(2n) & SO(2n+1) & USp(2n)\\
\hline
\end{array}
$$
This list implies that
an $\wt{\mbox{O}p}^-$-plane can be interpreted as
an O$p^-$-plane with a half D$p$-brane stuck on it.

For $p\geq6$, however, we have no ${\bf Z}_2$ torsion associated with
R-R fields. This seems to imply that there are only two kinds of
orientifold planes O$p^\pm$.
At first sight, it seems to be
possible to construct an $\wt{\mbox{O}p}^-$-plane by
considering a half D$p$-brane stuck on an O$p^-$-plane.
However, there are arguments from gauge theories on probe D-branes
which exclude the existence of $\wt{\mbox{O}6}^-$-planes and 
$\wt{\mbox{O}7}^-$-planes.
First, let us consider an $\wt{\mbox{O}7}^-$-plane.
The theory on one probe D3-brane near the O$7^-$-plane
is ${\cal N}=2$ $SU(2)$ gauge theory.
The number of complex fermion doublets is twice the number of background
D7-branes.
If there is a stuck D7-brane on
the O$7^-$-plane, which is counted as $1/2$ D-brane,
the number of fermion doublets is odd and the field theory suffers
from
the Witten's anomaly \cite{su2anomaly}.
Therefore, a stuck D$7$-brane on an O$7^-$-plane, i.e. 
an $\wt{\mbox{O}7}^-$-plane, is not allowed.

For an $\wt{\mbox{O}6}^-$-plane, a relevant anomaly is the parity
 anomaly \cite{Redlich,3dim} on a probe D2-brane.
In three dimensional $SU(2)$ gauge theory,
a fermion $1$-loop induces the following Chern-Simons term
for each complex fermion doublet.
\begin{equation}
\Gamma
=\frac{1}{8\pi}\sign(m)\int_3\tr(A\wedge dA
+\frac{2}{3}A\wedge A\wedge A)
=\frac{1}{8\pi}\sign(m)\int_4\tr(F\wedge F),
\label{pa}
\end{equation}
where $m$ is a real fermion mass
and the second integral is taken over a
four-manifold whose boundary is the three-dimensional spacetime.
By a large gauge transformation which changes the instanton number
in the four-manifold by one, the effective action is changed by
\begin{equation}
\Delta\Gamma=2\pi\frac{\sign(m)}{2}.
\end{equation}
If the number of fermion doublets is odd,
this large gauge transformation is broken regardless of the fermion masses.
This seems to prohibit the existence of a stuck D$6$-brane on an
O$6^-$-plane, i.e. an $\wt{\mbox{O}6}^-$-plane.

From these facts, one may jump at the conclusion that 
$\wt{\mbox{O}p}^-$-planes for $p\geq6$ do not exist.
As we will see later, this observation is partially correct.
The situation, however, is not so simple.
For example, considering T-duality, we immediately come up against
a puzzle.
That is, because there are both O$5$-planes and $\wt{\mbox{O}5}^-$-planes,
it seems possible to construct an $\wt{\mbox{O}6}^-$-plane by
T-duality from a pair of O$5$ and $\wt{\mbox{O}5}^-$.
If this is true, we would also be able to construct an
$\wt{\mbox{O}7}^-$-plane
as a T-dual configuration of a pair of O$6$ and $\wt{\mbox{O}6}^-$.

In this paper, we analyze
consistency of orientifold configurations and
propose a resolution of this puzzle.
In particular, we will show that $\wt{\mbox{O}6}^-$-planes do exist,
while $\wt{\mbox{O}7}^-$-planes and $\wt{\mbox{O}8}^-$-planes do not.
Interestingly, the absence of $\wt{\mbox{O}7}^-$ and $\wt{\mbox{O}8}^-$
is related to the cancellation of ${\bf Z}_2$ charges of
non-BPS D7-branes and non-BPS D8-branes in type I string theory.
 Actually, some arguments related to
 non-BPS D-branes turn out to be useful in our analysis.

We mainly focus our attention on O$p^-$ and $\wt{\mbox{O}p}^-$,
though we also obtain several new results concerned with
the other types of orientifold planes.

The outline of this paper is as follows.
In section {\ref{Wil.sec}} we analyze
Wilson lines in toroidally compactified type I string theory,
and give some rules to obtain consistent Wilson lines.
The constraints for the consistent configurations with
$\wt{\mbox{O}p}^-$-planes
are related to these rules by T-duality.
We will show in section \ref{tdual.sec} that
the results in section {\ref{Wil.sec}} are
actually consistent under T-duality, with paying attention
to the puzzle mentioned above.
Section {\ref{disc.sec}} is devoted to develop
general arguments about orientifolds.
We interpret discrete torsions of orientifolds
by using spherical D-branes and NS-branes wrapped around orientifold planes.
In section \ref{non-bps.sec} we will make
several comments about the relations between
non-BPS D-branes and orientifolds.
One of the observation given in section \ref{disc.sec} and \ref{non-bps.sec}
is that we can continuously take off stuck D$p$-branes
from $\wt{\mbox{O}p}$-planes in some configurations.
We further confirm this phenomenon in
section {\ref{transfer.sec}} from the viewpoint of
the corresponding Wilson lines in type I string theory.

%%%%%%%%%%%%%%%%%%%%%%%%%%%%%%%%%%%%%%%%%%%%%%%%%%%%%%%%%%%%%%%%%%%%%%%%
\section{Wilson Lines in Type I String Theory}\label{Wil.sec}

In this section, we will analyze
Wilson lines in toroidally compactified type I string theory.
Before discussing Wilson lines,
we have to clarify the global structure of the gauge group in
type I string theory.
Locally, gauge group is isomorphic to $Spin(32)$.
The center of $Spin(32)$ is ${\bf Z}_2^L\times{\bf Z}_2^R$ and
the vector, the spinor and the conjugate spinor representations carry
the central charge
$(-,-)$, $(-,+)$ and $(+,-)$ respectively.
All fields in perturbative type I string theory
are neutral with respect to this center
because an open string always has two end points belonging to the
vector representation.
Therefore, the perturbative gauge group is
$Spin(32)/({\bf Z}_2^L\times{\bf Z}_2^R)=SO(32)/{\bf Z}_2$.
Furthermore,  we can enlarge this gauge group to $O(32)/{\bf Z}_2$,
since transformations of determinant $-1$ are also symmetries in
type I perturbative string theory.
Note that
elements of $O(32)$  with determinant $-1$ 
exchange representations with central charge $(-,+)$ and those with
$(+,-)$.
The perturbative spectrum of type I string theory
is invariant under this operation.

Non-perturbatively, however, the situation becomes different.
Recently, it was found that type I 
string theory has stable non-BPS D-instantons
and stable non-BPS D-particles in its spectrum\cite{nonbps,WitKO}.
The existence of these objects suggests that the non-perturbative
gauge group of type I string theory is $Spin(32)/{\bf Z}_2^R$,
as expected from the type I-heterotic duality.
Actually, the non-BPS D-particles belong to
the spinor representation with central charge $(-,+)$,
which is not invariant under ${\bf Z}_2^L$.

Let us argue Wilson lines of this non-perturbative gauge group.
We will discuss only Wilson lines with vector structure.
This implies that dual orientifolds contain only O$p^-$ and
$\wt{\mbox{O}p}^-$
\cite{novector}. O$p^+$ and $\wt{\mbox{O}p}^+$ are not considered in
this section.
Instead, thanks to the vector structure, we can diagonalize all
Wilson lines
simultaneously on the vector representation.
Furthermore, to make our argument simple,
we assume all diagonal elements of Wilson lines are $+1$ or $-1$.
The incorporation of diagonal elements of generic values
does not change the arguments below.

First, let us consider type I string theory
compactified on ${\bf S}^1$ with a Wilson line $g_1$.
Here the subscript of $g$ denotes the
compactified direction.
Because $g_1$ has to be an element of $Spin(32)/{\bf Z}_2^R$,
the following condition should hold.
\begin{quote}
\underline{Condition A}\\
 \it{The number of $-1$ components of the Wilson line is even.}
\end{quote}
For example, the following Wilson line is not allowed.
\begin{equation}
g_1=\diag(-1,+1,(+1)^{30}).
\label{S1WL}
\end{equation}
If we allow this Wilson line, it causes an inconsistency that 
the chirality of the non-BPS D-particle flips 
when it goes around the compactified direction, because
$g_1$ is represented by $\Gamma \Gamma^1$ on the spinor representation.
Here $\Gamma^a$ are gamma matrices of $SO(32)$ and
$\Gamma=\Gamma^1\cdots\Gamma^{32}$. 
It is well known that Wilson lines correspond to the positions
of D-branes in the T-dual picture.
Therefore, the T-dual configuration of the Wilson line (\ref{S1WL})
has two $\wt{\mbox{O}8}^-$-planes.
So forbiddance of the Wilson line (\ref{S1WL}) implies that
$\wt{\mbox{O}8}^-$-planes in the T-dual picture are not allowed.

The Wilson line (\ref{S1WL}) corresponds to the non-trivial element of 
$\pi_0(O(32))={\bf Z}_2$
and is studied in many works in connection with non-BPS D8-branes.
Although it would be possible to construct any number of eight-branes
perturbatively, 
the ${\bf Z}_2$ charge should be cancelled on ${\bf S}^1$
in order to be compatible with
the existence of non-BPS D-particles, as explained above.
We will reconsider this point in section \ref{non-bps.sec}.

Let us move on to ${\bf T}^2$ compactification of
type I string theory and consider the following Wilson lines.
\begin{eqnarray}
g_1&=&\diag(-1,+1,-1,+1,(+1)^{28}),\nonumber\\
g_2&=&\diag(-1,-1,+1,+1,(+1)^{28}).
\label{T2WL}
\end{eqnarray}
If these Wilson lines are possible,
it would give four $\wt{\mbox{O}7}^-$-planes as a T-dual configuration.
However, we can again show this is not allowed as follows.

In general, Wilson lines for any two directions
should commute. More precisely, they should commute on the
representation of any matter field in the theory.
In type I string theory, since the non-BPS D-particle
 belongs to the spinor representation,
 $g_1$ and $g_2$ should commute on the spinor representation.
The spinor representations of these group elements are
\begin{equation}
g_1=\Gamma^1\Gamma^3,\quad
g_2=\Gamma^1\Gamma^2.
\end{equation}
Clearly, these two matrices do not commute with each other.
(Instead, they anti-commute.)
Therefore, these Wilson lines are not allowed and we cannot
make $\wt{\mbox{O}7}^-$-planes as a T-dual configuration.

Topologically, the Wilson lines (\ref{T2WL}) correspond to a
non-trivial element of homotopy group $\pi_1(O(32))={\bf Z}_2$
which corresponds to a non-BPS D7-brane.  The argument above
implies that the ${\bf Z}_2$ charge carried by non-BPS D7-branes
must be cancelled on ${\bf T}^2$, as in the case of non-BPS D8-branes. 
See section \ref{non-bps.sec} for more details.

In order for any two Wilson lines to commute,
we must impose the following condition in addition to the condition A.
\begin{quote}
\underline{Condition B}\\
\it{For any two $g_i$ and $g_j$, the number of components which
are $-1$ for both $g_i$ and $g_j$ is even.}
\end{quote}
This statement is obtained by using the fact
that the spinor representation of $g_i$ is given as
a product of all gamma matrices having indices
of $-1$ components.
(Thanks to Condition A, $\Gamma$ insertion is not necessary.)

Let us proceed to the case of ${\bf T}^3$ compactification.
We can make the following Wilson lines.
\begin{eqnarray}
g_1&=&\diag(-1,+1,-1,+1,-1,+1,-1,+1,(+1)^{24}),\nonumber\\
g_2&=&\diag(-1,-1,+1,+1,-1,-1,+1,+1,(+1)^{24}),\nonumber\\
g_3&=&\diag(-1,-1,-1,-1,+1,+1,+1,+1,(+1)^{24}).
\label{T3WL}
\end{eqnarray}
These Wilson lines
satisfy the two conditions A and B
and hence are compatible with the non-perturbative gauge group.
We claim that these Wilson lines are really possible and
the T-dual configuration with eight $\wt{\mbox{O}6}^-$-planes is as well.
This seems to be inconsistent with the argument of
discrete torsions.
We will discuss this point in the next section.

Although we can construct Wilson lines
dual to eight O$6^-$ or eight $\wt{\mbox{O}6}^-$,
there are no Wilson lines dual to mixed configurations
containing both O$6^-$ and $\wt{\mbox{O}6}^-$.
The proof is as follows.
On the orientifold ${\bf T}^3/{\bf Z}_2$, there are eight O$6$-planes.
We normalize their positions on the
${\bf T}^3/{\bf Z}_2$ as $(\pm1,\pm1,\pm1)$,
i.e., as apexes of a cube with sides $2$.
The two conditions above demand two adjacent apexes to be the same
type of
orientifold planes.
And this results that all apexes must be the same type of O$6$-planes.
At first sight, this may seem strange because
this implies there is some correlation among eight O$6$-planes.
In the next section, we will see
this is explained in a natural way in the dual orientifold picture.

\begin{figure}[htb]
\begin{center}
  \leavevmode
  \epsfxsize=135mm
  \epsfbox{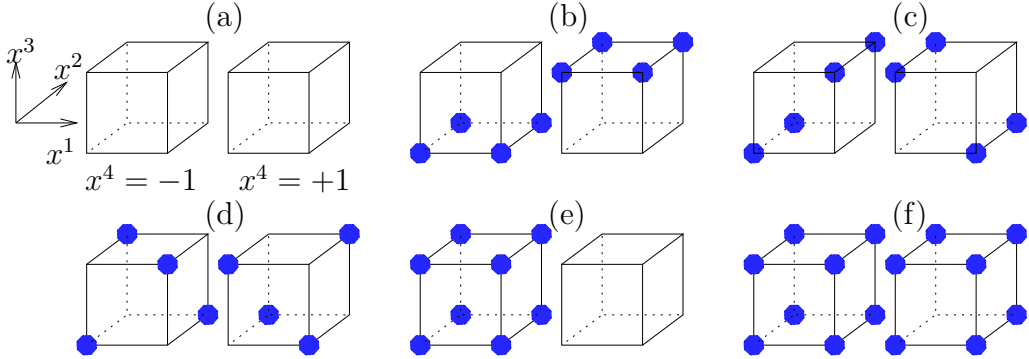}
    \put(-312,125){(a)}
    \put(-182,125){(b)}
    \put(-52,125){(c)}
    \put(-312,50){(d)}
    \put(-182,50){(e)}
    \put(-52,50){(f)}
    \put(-372,73){$x^1$}
    \put(-369,105){$x^2$}
    \put(-387,111){$x^3$}
    \put(-357,63){$x^4=-1$}
    \put(-299,63){$x^4=+1$}
  \caption{\small
  Each possible configuration of O$5$-planes is the same with
  one of these up to rotation.
  The cubes spread in the $x^1$,$x^2$,$x^3$ 
  directions and left (right) ones are located at $x^4=-1$ ($+1$).
  O$5$-planes are located at the apexes of the cubes, and
  the bullets represent $\wt{\mbox{O}5}^-$-planes.
  We have 
  1 (a)-type, 12 (b)-type, 8 (c)-type, 2 (d)-type, 8 (e)-type and 1
  (f)-type configurations.
  (b), (c), (d) and (e) are equivalent to each other via modular
  transformations.}
  \label{zu:O5}
\end{center}
\end{figure}

For a ${\bf T}^4/{\bf Z}_2$ orientifold,
$16$ O$5$-planes are put on each apex of a four-dimensional cube
and the two conditions above demand the number
of $\wt{\mbox{O}5}^-$ on each (two-dimensional) face should be
even: $0$, $2$ or $4$.
Let $\alpha$ denote an apex of the four-dimensional cube and
$P_i\alpha$ ($i=1,2,3,4$) denote its reflection along the $i$-th axis.
If we introduce $\sigma(\alpha)$
which is $+1$ ($-1$) for $\alpha$ corresponding to O$5^-$
($\wt{\mbox{O}5}^-$),
the conditions above are equivalent
 to the statement that the following value
does not depend on the apex $\alpha$.
\begin{equation}
c_i=\sigma(P_i\alpha)\sigma(\alpha).
\end{equation}
Therefore, general solution for the conditions is represented as
follows.
\begin{eqnarray}
\sigma(P_i\alpha)&=&c_i\sigma(\alpha),\nonumber\\
\sigma(P_iP_j\alpha)&=&c_ic_j\sigma(\alpha),\quad (i<j),\nonumber\\
\sigma(P_iP_jP_k\alpha)&=&c_ic_jc_k\sigma(\alpha),\quad
(i<j<k),\nonumber\\
\sigma(P_1P_2P_3P_4\alpha)&=&c_1c_2c_3c_4\sigma(\alpha).
\end{eqnarray}
where $\alpha$ in these equations is fixed.
Because we can freely choose five variables, $c_i$ and
$\sigma(\alpha)$,
there are $32$ allowed configurations.
It is easy to construct these $32$ possible configurations explicitly,
and we find that 
the number of $\wt{\mbox{O}5}^-$ is restricted to be
one of $0$, $8$ and $16$.
(Fig.\ref{zu:O5})

On a ${\bf T}^n/{\bf Z}_2$ orientifold, there are $2^n$ O$(9-n)$-planes.
They are labeled by $n$-dimensional vectors with components $\pm1$.
Let us define a subset of these O$(9-n)$-planes
whose labels have $(n-3)$ or less $+1$ components.
This subset contains $\sum_{k=0}^{n-3}{}_nC_k$ O$(9-n)$-planes,
and we can freely choose the types of them.
The types of the other O-planes
are uniquely determined by the two conditions.
Therefore, the number of possible configurations is
\begin{equation}
N=2^{\sum_{k=0}^{n-3}{}_nC_k}=2^{2^n-(n^2+n+2)/2}.
\end{equation}
When $n\geq6$, the number of orientifold planes is larger than $32$
and some of $N$ configurations should contain anti D-branes
so that the total R-R charge is cancelled.
Because we have not used the supersymmetry, our argument is applicable
even in such cases.

We have obtained
two necessary conditions for the Wilson lines to be allowed.
We expect that these conditions are sufficient to
select possible Wilson lines.
However, it is not a simple task to prove this statement.
In the next section, we will discuss T-duality among
orientifold planes with different dimensions.
We will show the argument is very consistent with the
two conditions we obtained in this section.
It strongly suggests the conditions are sufficient.

%%%%%%%%%%%%%%%%%%%%%%%%%%%%%%%%%%%%%%%%%%%%%%%%%%%%%%%%%%%%%%%%%%%%%%%%%%
\section{T-duality of Orientifold Planes}\label{tdual.sec}

Having analyzed Wilson lines on torus in type I string theory,
some questions arise.
First, we saw that the conditions for the possible
Wilson lines in type I string theory suggests the existence
of both O$6^-$ and $\wt{\mbox{O}6}^-$.
Is this consistent with the analysis of the discrete torsion
and the anomaly mentioned in section \ref{intro.sec}?
Next, we saw that we cannot freely choose the types of
orientifold planes in a compact orientifold.
How does this correlation among orientifold planes
explained?
Finally, how is it prohibited to make
$\wt{\mbox{O}7}^-$ or $\wt{\mbox{O}8}^-$
by T-duality?

To answer these questions, let us consider T-duality between an
$\wt{\mbox{O}6}^-$-plane and a pair of O$5^-$ and $\wt{\mbox{O}5}^-$.
\begin{equation}
\mbox{O}5^-+\wt{\mbox{O}5}^- \leftrightarrow \wt{\mbox{O}6}^-.
\label{T-dual}
\end{equation}
Here we assume the positions of the O$5^-$-plane and
$\wt{\mbox{O}5}^-$-plane
to be $(x^1,x^2,x^3,x^4)=(0,0,0,0)$ and $(0,0,0,a)$ respectively.
The ${\bf Z}_2$ actions for these orientifold planes are
\begin{equation}
\Omega : (x^1,x^2,x^3,x^4)\rightarrow(-x^1,-x^2,-x^3,-x^4),
\end{equation}
\begin{equation}
\wt{\Omega} : (x^1,x^2,x^3,x^4)\rightarrow(-x^1,-x^2,-x^3,2a-x^4),
\end{equation}
respectively. 
The composition of these transformations is a shift along $x^4$ by $2a$,
\begin{equation}
\wt{\Omega}\Omega : (x^1,x^2,x^3,x^4)\rightarrow(x^1,x^2,x^3,x^4+2a).
\label{parat}
\end{equation}
This implies $x^4$ direction is compactified with a period $2a$.
The T-duality is taken along this direction.

The orientifold flip acts on R-R fields non-trivially.
Especially, $\Omega$ and $\wt{\Omega}$ change 
the sign of the axion field $C/2\pi$.
Taking account of the freedom of shifting by an integer,
we obtain the following transformation low.
\begin{equation}
\Omega : C(x^4)/2\pi\rightarrow n-C(-x^4)/2\pi,\quad
\wt{\Omega} : C(x^4)/2\pi\rightarrow n'-C(2a-x^4)/2\pi.
\label{axionflip}
\end{equation}
(The dependence on the other coordinates is omitted.)
{}From (\ref{axionflip}), the value of the axion field on top
of an orientifold plane should be a half integer.
Up to an integral shift, we have two physically distinct
cases: $C/2\pi\in{\bf Z}$ and $C/2\pi\in {\bf Z}+1/2$.
In fact, this value represents the discrete torsion
associated with the orientifold plane.
Therefore, the integral and half odd integral axion field
correspond to O$5^-$ and $\wt{\mbox{O}5}^-$ respectively.
In our case this fact implies
\begin{equation}
n\in2{\bf Z},\quad
n'\in2{\bf Z}+1.
\end{equation}
Now, let us see how the axion field is transformed under the
parallel transport (\ref{parat}).
The composition of (\ref{axionflip}) is
\begin{equation}
\wt{\Omega}\Omega : C(x^4)/2\pi\rightarrow C(x^4+2a)/2\pi-(n'-n).
\end{equation}
Notice that the change of the axion field $(n'-n)$ is always an
odd integer
and does not vanish.
This is represented by the following equation.
\begin{equation}
\frac{1}{2\pi}\oint dC=n'-n,
\end{equation}
where the integral is taken along the ${\bf S}^1$.
Under the T-duality along the $x^4$ direction,
the R-R one-form field strength on the left hand side is
transformed into R-R zero-form field strength, which is electric-magnetic
dual to the ten-form field strength of the R-R nine-form field.
Namely, the configuration with different kinds of O$5$-planes
are transformed into `massive' type IIA theory
with odd cosmological constant.\footnote{
In this paper, the R-R zero-form field strength is often called
`the cosmological constant', and normalized to be an integer.
Actually,
it is the square root of the cosmological constant in a
conventional sense,
since it induces the
cosmological term $S\sim\int d^{10}x\sqrt{-g} (F_0)^2$.
We hope this expression will not lead any confusion.
}

The generalization of the argument above
to include the case of a pair of the same kind of O$5$ is
straightforward and
the generalized statement is as follows.
\begin{quote}
\it{A pair of the same kind of O$5$-planes is transformed into
O$6^-$ in the background with even cosmological constant,
while a pair of different kinds of O$5$-planes is mapped into
$\wt{\mbox{O}6}^-$ in the odd background cosmological constant.}
\end{quote}

Although both O$6^-$ and $\wt{\mbox{O}6}^-$ are possible,
once the background cosmological constant is fixed,
one of them is automatically chosen.
Namely, O$6^-$ ($\wt{\mbox{O}6}^-$) are possible only in the
background with
even (odd) cosmological constant.
This is consistent with the absence of ${\bf Z}_2$ torsion.
Instead of ${\bf Z}_2$ torsion, we have a non-trivial cohomology group
$H^0({\bf RP}^2,{\bf Z})={\bf Z}$ associated with the
R-R 0-form field strength.
Therefore, the element of this cohomology group 
gives an extra charge of the O$6$-plane,
which is identified with the background cosmological constant.
This naturally gives the reason why all O$6$-planes
in the ${\bf T}^3/{\bf Z}_2$ orientifold
 belong to the same type.
Because the background cosmological constant fixes the type of
O$6$-planes,
 all O$6$-planes in a BPS configuration have to be the same type.

Taking account of the relation between the cosmological constant and
the type of O$6$-planes, 
the parity anomaly problem mentioned in the section \ref{intro.sec}
can be solved.
The action of massive D2-brane contains the following term
\cite{massived2}.
\begin{equation}
S=\frac{\Lambda}{8\pi}\int \tr(A\wedge dA
+\frac{2}{3}A\wedge A\wedge A),
\label{massiveCS}
\end{equation}
where $\Lambda$ is the integral background cosmological constant. 
This term supply the bare Chern-Simons term in the bare action
and combined with the one-loop effect of world-volume fermions
(\ref{pa}),
the invariance under large gauge transformations
is recovered.\footnote{%
The trace in eq.(\ref{massiveCS}) is taken on the $\bf 2k$ representation of
the $USp(2k)$ gauge group.
If the D2-branes are moved away from the orientifold plane,
the gauge group is broken to $U(k)$.
If we use generators for the fundamental representation of this $U(k)$,
the coefficient in eq.(\ref{massiveCS}) is doubled because $\bf2k$ is decomposed
into ${\bf k}+\ol{\bf k}$ representation of $U(k)$.
Therefore, the argument above does not conflict with the existence of
massive D2-branes in odd $\Lambda$ backgrounds.}

Now, we have shown that both kinds of O$6$-planes are possible
and solved the puzzle about T-duality between O$5$-planes and
O$6$-planes.
The T-duality between O$6$ and O$7$ should be considered next.
One may suspect that we can have
a pair of O$6^-$ and $\wt{\mbox{O}6}^-$ in some non-BPS configuration
with varying cosmological constant.
If so,
is it possible to obtain $\wt{\mbox{O}7}^-$-planes via T-duality
transformation,
\begin{equation}
\mbox{O}6^-+\wt{\mbox{O}6}^-\leftrightarrow \wt{\mbox{O}7}^-~?
\end{equation}
As explained in section \ref{intro.sec},
The existence of the $\wt{\mbox{O}7}^-$-plane
causes Witten's anomaly on the D3-brane 
probe field theory.
Furthermore, half D-brane charge due to the stuck D7-brane on
the O$7^-$-plane
implies the non-integral monodromy associated with the axion field.
This is physically unacceptable.

The resolution of this puzzle is as follows.
If we wish to consider the configurations with 
both O$6^-$ and $\wt{\mbox{O}6}^-$, the two regions should be divided
by odd number of D8-branes
playing a role of domain walls.
For example, let us assume $\wt{\mbox{O}6}^-$ is enclosed spherically
by a D8-brane.
(This D8-brane is `spherical' on the covering space.
After ${\bf Z}_2$ orientifolding it becomes ${\bf RP}^2$.)
Actually, the non-trivial twisted homology group 
$H_2({\bf RP}^2,\wt{\bf Z})={\bf Z}$ shows such a wrapping is possible.
Spherical D-brane configurations
are considered in \cite{KT} and \cite{qball}
in the context of matrix theory
and in Myers\cite{myers} showes that such configurations are stabilized
in the non-zero R-R field strength background.
(Of course, in our case, it does not have to be stabilized at all.)
Recently, such configurations are used
to analyze ${\cal N}=1$ gauge theory \cite{ps,ar}.
As discussed in \cite{ar},
in order to be consistent with the results in
field theory,
the D$p$-brane charge carried by a spherical D$(p+2)$-brane
enclosing an orientifold $p$-plane
should be shifted by half, which is induced by the shifted
flux quantization condition,
\begin{equation}
\frac{1}{2\pi}\oint_{{\bf RP}^2}f_2\in{\bf Z}+\frac{1}{2}.
\label{quant}
\end{equation}
Here $f_2$ is the field strength of the gauge field on
the spherical D$(p+2)$-brane.
This is also suggested by the argument of stringy anomaly
 in \cite{anomalies}. 
We will show the equation (\ref{quant})
in section \ref{non-bps.sec}, using an argument
of tachyon condensation.
Therefore, the total amount of the D6-brane charge of
$\wt{\mbox{O}6}^-$ enclosed by a spherical D8-brane
is the same with that of O$6^-$ (up to integer).
Therefore, when the spherical D8-brane shrinks to a point,
the pair of O$6^-$ and $\wt{\mbox{O}6}^-$ reduces to
a pair of O$6^-$-planes!
This results that the T-dual of the O$6$ configurations are
always not $\wt{\mbox{O}7}^-$ but O$7^-$.

Up to now, we have seen that
our results on $\wt{\mbox{O}6}^-$ and $\wt{\mbox{O}7}^-$
in the last section is also explained in the dual orientifold picture
by using the discrete torsions.
In fact, we can show the two conditions A and B are equivalent
to the requirement
that the discrete torsions on the orientifold should be well-defined.
On a ${\bf T}^n/{\bf Z}_2$ orientifold with $n\geq4$,
a section with three coordinates $x^i$, $x^j$ and $x^k$ fixed
is ${\bf T}^{n-3}$.
The integral of the R-R $(n-3)$-form field strength on this section
should not depend on the position $(x^i,x^j,x^k)$.
For example, the integral at $(+1,+1,+1)$ and that at $(+1,+1,-1)$
must be the same.
These two integrals are nothing but the sums of the
discrete torsions of O$(9-n)$-planes on the sections.
Therefore, the requirement above implies the sum of
${\bf Z}_2$ discrete torsions of O$(9-n)$-planes on a section with $x^i$
and $x^j$ fixed
should be zero.
This is equivalent to the conditions we obtained in section
\ref{Wil.sec}.

Before ending this section,
we would like to give one more consistency check.
Let us consider the Wilson lines (\ref{T3WL})
and take a T-duality transformation along the $x^3$ direction.
This system consists of two O$8^-$-planes with
four D8-branes on each of them.
(We omit last $24$ components in (\ref{T3WL}) because they don't
affect the following arguments.)
The gauge group on each world-volume of four D$8$-branes is (locally)
$Spin(4)$ and each has the following Wilson lines.
\begin{eqnarray}
g_1&=&\diag(-1,+1,-1,+1),\nonumber\\
g_2&=&\diag(-1,-1,+1,+1).
\label{O8WL}
\end{eqnarray}
This has the same structure with the Wilson lines (\ref{T2WL}).
Recall that the Wilson lines (\ref{T2WL}) are
unacceptable, since $g_1$ and $g_2$ in (\ref{T2WL})
do not commute on the spinor representation, to which
non-BPS D-particles belong.
Now we are in face of the same problem.
Because (BPS) D-particles stuck
on one of the O$8^-$-planes
belong to spinor representation
of the $Spin(4)$ gauge group on the world-volume of four D$8$-branes,
it seems that the Wilson lines (\ref{O8WL}) are not acceptable
as the Wilson lines (\ref{T2WL}) are not.
If so, in addition to the two conditions we have already discussed,
we have to introduce some stronger constraint.
Can it be true?
If we should introduce such a constraint, a lot of configurations,
including (\ref{T3WL}), would be prohibited
and the arguments so far become almost meaningless.

Before solving this problem, let us restate the 
reason why the Wilson lines (\ref{O8WL}) lead to the problem
in a more convenient form.
Let us consider closed world-line $C$ of a stuck D-particle on the
world-volume of four D8-branes.
If $C$ wrapped around the
$x^2$ direction, the contribution of the $Spin(4)$ gauge field $A^a$ to the
partition function
of the D-particle is
\begin{equation}
Z_A=\tr\left(g_2P\exp\int_CA^aT_a\right).
\label{z2}
\end{equation}
where $T_a$ is a generator
of $Spin(4)$ gauge group in the spinor representation.
The insertion of $g_2$ is necessary due to the Wilson line (\ref{O8WL})
along the $x^2$-direction.
If we move the cycle $C$ to go around the cycle along $x^1$,
the partition function changes as follows
\begin{equation}
Z_A'=\tr\left(g_1g_2g_1^{-1}P\exp\int_CA^aT_a\right).
\label{z3}
\end{equation}
Because $g_1g_2g_1^{-1}$ is equal to $-g_2$
In the spinor representation,
$Z_A$ and $Z_A'$ do not coincide.
This is against the singlevaluedness of the partition function.

A key saving us from this situation is the fact that
the dual O$6$-plane configuration
of (\ref{T3WL}) always has odd cosmological constant.
It implies that in the O$8$-plane configuration
the R-R two-form field strength $F_{12}$
is non-zero and the total flux is odd integer.
Taking account of the fractional charge of a stuck D-particle,
the contribution of R-R $1$-form field to the partition function is
\begin{equation}
Z_{RR}=\exp\left(\frac{i}{2}\oint_CC_1\right).
\end{equation}
If we move the path $C$ around $x^1$, it sweeps out the $x^1$-$x^2$ torus, and
the partition function changes to $Z_{RR}'$ satisfying
\begin{equation}
Z'_{RR}/Z_{RR}=\exp\left(\frac{i}{2}\oint_{{\bf T}^2}dC_1\right).
\label{zrrzrr}
\end{equation}
Because total flux is odd integer,
the left hand side of (\ref{zrrzrr}) is $-1$.
This correctly cancels the contribution of the Wilson lines
(\ref{O8WL}).
In other words, the gauge group associated with stuck D-particles
is not $Spin(4)\times U(1)$ but
\begin{equation}
G=(Spin(4)\times U(1))/{\bf Z}_2.
\label{gond8}
\end{equation}
The Wilson lines (\ref{O8WL}) are not well-defined as a gauge bundle
for $Spin(4)$
but they are for the gauge group (\ref{gond8}) if they are accompanied
by appropriate R-R gauge flux.

Note that this trick is not applicable to the Wilson lines
(\ref{T2WL})
because there is no R-R gauge field coupling to non-BPS D-particles in
type I string theory.

%%%%%%%%%%%%%%%%%%%%%%%%%%%%%%%%%%%%%%%%%%%%%%%%%%%%%%%%%%%%%%
%%%%%%%%%%%%%%%%%%%%%%%%%%%%%%%%%%%%%%%%%%%%%%%%%%%%%%%%%%%%%%
\section{Discrete Torsions and Spherical Branes}\label{disc.sec}

In the previous section,
we observed that an $\wt{\mbox{O}6}^-$-plane enclosed by
a spherical D8-brane is equivalent to an O$6^-$-plane.
We can easily generalize this argument to the cases including
other types of orientifold planes.

Our first claim is as follows.
\begin{quote}
\it{
We can change O$p$ to $\wt{\mbox{O}p}$ by
wrapping it spherically with a D$(p+2)$-brane,
and vice versa.}
\end{quote}
This statement is meaningful for O$p$-planes of
$p\le 6$.
Note that we have a non-trivial twisted 2-cycle ${\bf RP}^2$,
on which the D$(p+2)$-brane is wrapped,
as a submanifold of the ${\bf RP}^{8-p}$ enclosing the O$p$-plane,
since we have
\begin{eqnarray}
H_2({{\bf RP}^{8-p}},\wt{\bf Z})=
\left\{
\begin{array}{cc}
{\bf Z}_2 &  p<6,\\
{\bf Z} &  p=6.
\end{array}
\right.
\end{eqnarray}

To confirm this claim,
we have to show the wrapped D$(p+2)$-brane changes the discrete
torsion of
orientifold $p$-plane associated with the R-R $(5-p)$-form field.
We can prove this in a similar way as in
\cite{WittenBaryon}, in which arguments
 for O$3$-planes with spherical five-branes
 are given.
Let us consider a spherical D($p+2$)-brane with
a radius $r_0$ wrapped around an O$p$-plane.
The discrete torsion is defined by
\begin{equation}
I(r)=\frac{1}{2\pi}\oint_{{\bf RP}^{5-p}}C_{5-p}\quad \mod1,
\end{equation}
where $r$ is a radius of the ${\bf RP}^{5-p}$.
If $r$ is much larger than $r_0$,
this integral gives the discrete torsion of the
whole system of the O$p$-plane
and the spherical D($p+2$)-brane,
while in the $r\rightarrow0$ limit,
it gives that of only the O$p$-plane.

\begin{figure}[htb]
\begin{center}
  \leavevmode
  \begin{picture}(100,125)
  \epsfxsize=50mm
    \put(-20,0){\epsfbox{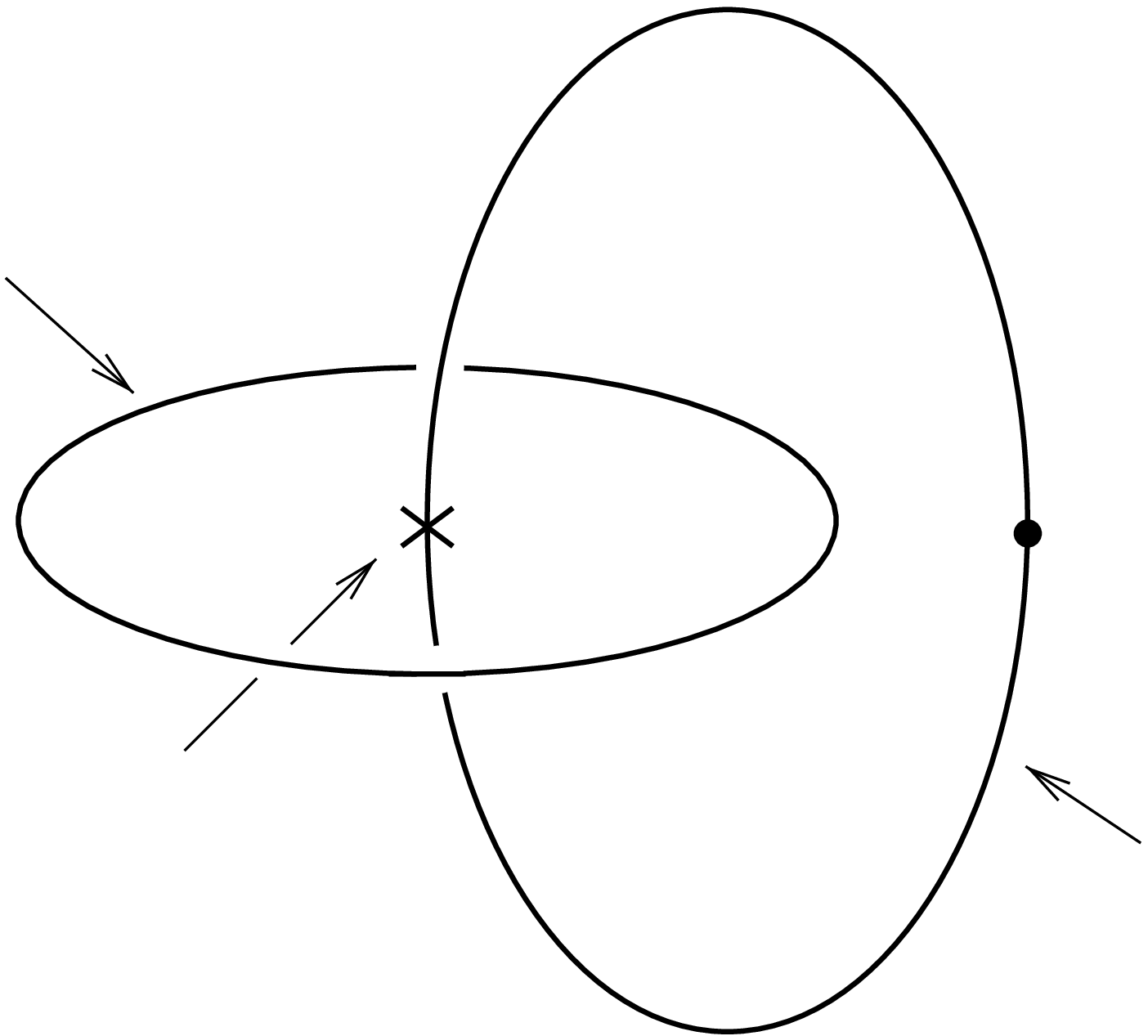}}
    \put(-15,25){O$p$}
    \put(112,59){$\infty$}
    \put(-33,95){${\bf S}^2$}
    \put(124,13){${\bf S}^{6-p}$}
  \end{picture}
\end{center}
  \caption{}
  \label{zu:link}
\end{figure}

In the covering space,
the D($p+2$)-brane is wrapped on ${\bf S}^{2}$
with radius $r_0$.
Let us think about a process in which $r$ changes from $0$ to
infinity.
In this process, ${\bf S}^{5-p}$ sweeps a $(6-p)$-dimensional
manifold.
If we compactify this manifold by adding a point at infinity,
it becomes ${\bf S}^{6-p}$ as depicted in Fig.\ref{zu:link}.
Because this ${\bf S}^{6-p}$ and the ${\bf S}^2$, on which the
D$(p+2)$-brane is wrapped,
are linked with each other
in a ($9-p$)-dimensional space perpendicular to the ${\rm O}p$-plane,
the following integral picks up the R-R charge of the D$(p+2)$-brane.
\begin{equation}
\frac{1}{2\pi}\oint_{{\bf S}^{6-p}}dC_{5-p}=1.
\end{equation}
By Stokes' theorem, we obtain
\begin{equation}
2I(r=\infty)-2I(r=0)=1.
\end{equation}
(We need the factor $2$ in the left hand side
because we are discussing in the covering space.)
This implies the wrapped D$(p+2)$-brane changes the
discrete torsion.

Moreover, we have to show that the wrapped D$(p+2)$-brane
changes the R-R charge of the O$p$-plane correctly.
It should change the R-R charge of O$p^-$ by $1/2$ and
should not change that of O$p^+$.
Therefore, the R-R D$p$-brane charge which the wrapped D$(p+2)$-brane
carries should be
\begin{equation}
Q=\frac{1}{2}+I_{NS}\quad \mod1,
\label{Qd}
\end{equation}
where $I_{NS}$ is the discrete torsion of the O$p$-plane
associated with the NS-NS $B_2$ field,
\begin{equation}
I_{NS}=\frac{1}{2\pi}\oint_{{\bf RP}^2}B_2\quad \mod1.
\end{equation}
This is explained as follows.
The action on the D$(p+2)$-brane involves a term
\begin{equation}
\frac{1}{2\pi}\int_{p+3}(f_2+B_2)\wedge C_{p+1},
\label{daction}
\end{equation}
where $f_2$, $B_2$ and $C_{p+1}$ are the field strength of
the gauge field on the D$(p+2)$-brane, the NS-NS $2$-form field
and the R-R ($p+1$)-form field respectively.
We can easily see that the coupling between $B_2$ and $C_{p+1}$
reproduces the second term in (\ref{Qd}).
Taking account of
the shifted quantization condition (\ref{quant}),
we can understand that the $1/2$ term in (\ref{Qd})
is provided by the coupling $f_2\wedge C_{p+1}$.

Next, we examine O$p$-planes with $p\le 5$
enclosed by NS5-branes.
A similar argument as above implies that
if we wrap an NS5-brane on ${\bf RP}^{5-p}$ enclosing an O$p$-plane,
the discrete torsion associated with the NS-NS 2-form field
changes.
Therefore, it is reasonable to claim the following.
\begin{quote}
\it{An NS5-brane wrapped on ${\bf RP}^{5-p}$ changes an O$p^-$-plane
($\wt{\mbox{O}p}^-$-plane)
to an O$p^+$-plane ($\wt{\mbox{O}p}^+$-plane),
and vice versa.}
\end{quote}
The D$p$-brane charge of an O$p^+$-plane and that of
$\wt{\mbox{O}p}^+$-plane
are both $+2^{p-5}$. Hence the wrapped NS5-brane should carry
the R-R charge of
\begin{equation}
Q=2^{p-4}+I_R,
\label{Qns}
\end{equation}
where $I_R$ is the discrete torsion associated with $C_{5-p}$,
which is defined by
\begin{equation}
I_R=\frac{1}{2\pi}\oint_{{\bf RP}^{5-p}}C_{5-p}\quad \mod1.
\end{equation}

The $I_R$ term in (\ref{Qns}) is easily reproduced in the same way
as we did above for the wrapped D-brane.
The NS5-brane action has the following term.
\begin{equation}
\frac{1}{2\pi}\sum_p\int_6(h_{5-p}+C_{5-p})\wedge C_{p+1},
\label{nsaction}
\end{equation}
where $h_{5-p}$ are the field strengths of the gauge fields
on the NS5-brane.
For type IIA (IIB) NS5-branes, the sum is taken over $p=0,2,4$
($p=-1,1,3,5$).
The $p=3$ term is S-dual of (\ref{daction}) and other terms
are required by the consistency with the T-duality.
The second term in (\ref{Qns}) is
generated by the coupling between $C_{5-p}$ and $C_{p+1}$ in
(\ref{nsaction}).
The first term in (\ref{Qns}) is explained by supposing the flux
quantization
\begin{equation}
\frac{1}{2\pi}\oint_{{\bf RP}^{5-p}}h_{5-p}\in{\bf Z}+2^{p-4}.
\label{hquant}
\end{equation}
We do not know how to prove this quantization condition directly,
but we can see this condition is consistent with duality
in some specific cases.
For the case of $p=3$, as expected by S-duality,
this is the same condition as (\ref{quant}). The condition
for the case of $p=4$ is consistent with the analyses of
the M-theory lift of O$4$-planes given in \cite{Hori,Gimon}.

\begin{table}
  \begin{center}
  \begin{tabular}{ccc}
    & ${\bf RP}^2$ D$(p+2)$ & \\
    \hspace{2cm} O$p^-$ & $\longleftarrow\!\!\!-\!\!\!\longrightarrow$ & 
    \hspace{-1.8cm} $\wt{\mbox{O}p}^-$ \\
    &&\\
    ${\bf RP}^{5-p}$ NS5 $\Bigg\updownarrow$ & & 
    \quad$\Bigg\updownarrow$ ${\bf RP}^{5-p}$ NS5 \\
    &&\\
    \hspace{2cm} O$p^+$ & $\longleftarrow\!\!\!-\!\!\!\longrightarrow$ & 
    \hspace{-1.8cm} $\wt{\mbox{O}p}^+$ \\
    & ${\bf RP}^2$ D$(p+2)$ & \\
  \end{tabular}
  \end{center}
  \caption{\small
  The relations between the types of orientifold planes
  and the wrapped branes.}
\end{table}

In the case of O$0$-planes and O$1$-planes, we can consider two extra
${\bf Z}_2$ discrete
torsions which are associated with the R-R $(1-p)$-form field and the
NS-NS $6$-form field
respectively\cite{HananyKol}.
We can prove
the following statement
in the same way as above.
\begin{quote}
\it{For O$p$-planes with $p\leq1$,
we can change the discrete torsions with respect to the R-R
$(1-p)$-form field
and the NS-NS $6$-form field by wrapping a D$(p+6)$-brane on
${\bf RP}^6$
and a fundamental string on ${\bf RP}^{1-p}$ respectively.}
\end{quote}
O$2$-planes also have extra non-trivial cohomology group
$H^0({\bf RP}^6,{\bf Z})={\bf Z}$.
It is not torsion but an integral charge.
This charge is identified with the background cosmological constant
just as the $H^0$-cohomology for O$6$-planes.
In fact, for $p=2$, the wrapped D-brane is a D8-brane,
and this plays a role of a domain wall which divide the 
background into two regions with different cosmological constant.

%%%%%%%%%%%%%%%%%%%%%%%%%%%%%%%%%%%%%%%%%%%%%%%%%%%%%%%%%%%%%%%%%%%%%

\section{Non-BPS D-branes and Orientifold Planes}\label{non-bps.sec}

In this section, we argue some relations between
$\wt{\mbox{O}p}^-$-planes 
and stable non-BPS D-branes in type I string theory. We will
interpret the results in the previous sections
in terms of non-BPS D-branes.

To begin with, let us review stable
non-BPS D-branes in type I string
theory in brief. 
As shown in \cite{WitKO},
the D$p$-brane charges in type I string theory can be classified by 
the reduced K-theory group $\wt{KO}({\bf S}^{9-p})$,
or equivalently by the $(8-p)$th homotopy group
of the perturbative gauge group $O(32)$.
The non-trivial homotopy groups of $O(32)$ are given as follows,
\begin{eqnarray}
  &&\pi_7 (O(32)) = \pi_3 (O(32)) = \bf{Z}, \label{eq:homoto1} \\
  &&\pi_9 (O(32)) = \pi_8 (O(32)) = \pi_1 (O(32)) = \pi_0 (O(32))
  = {\bf Z}_2.
  \label{eq:homoto2}
\end{eqnarray}
The first line corresponds to BPS D1 and D5-brane charges and the second line 
corresponds to non-BPS D($-1$), D0, D7 and D8-brane ${\bf Z}_2$ charges,
respectively \cite{WitKO}.
Note that the non-BPS D7 and D8-branes in type I string theory
are known to have some subtleties in the arguments of the stability
\cite{FGLS,Sch}.
Namely, the open strings stretched between these objects and
background D9-branes create tachyonic modes which may cause
the instability. However, it will not affect our arguments
below, since
we only consider the topological structure of the system
which will remain after the tachyon condensation.

Let us consider type I string theory compactified on ${\bf S}^1$.
A non-BPS D8-brane can be constructed in the same way
as the construction of a non-BPS D-particle using a half D1-\AD1
\footnote{
In this paper, we count the number of D-branes after the
orientifold projection. Therefore, the gauge group on $n$ D1-branes
(or D9-branes)
in type I string theory is $O(2n)$. For $n=1/2$, the gauge group
is $O(1)={\bf Z}_2=\{\pm 1\}$.} pair \cite{nonbps}.
Consider a half D9-\AD9 pair wrapping the ${\bf S}^1$,
and assign the $-1$ Wilson line on the half D9-brane and
the $+1$ Wilson line on the half \AD9-brane. This system
is topologically equivalent to a non-BPS D8-brane.
Adding background 16 D9-branes, we obtain the following Wilson line
\begin{eqnarray}
g_1=\diag(-1,(+1)^{32}),~~ g_1'= +1,
\end{eqnarray}
where $g_1'$ denotes the Wilson line on the half \AD9-brane.
This is equivalent
to the Wilson line
\begin{eqnarray}
g_1=\diag(-1,+1,(+1)^{30}),
\label{WilA}
\end{eqnarray}
up to creation and
annihilation of D-brane - anti D-brane pairs.
This Wilson line actually picks up the non-trivial element of
$\pi_0(O(32))={\bf Z}_2$,
which corresponds to the non-BPS D8-brane charge as mentioned above.
We argued in section \ref{Wil.sec} that this Wilson line
is not allowed, since it is not compatible
with the existence of non-BPS D-particles. In other words,
we must have even number of non-BPS D8-branes which
are transverse to the
${\bf S}^1$. We will come back to this point later,
but now let us formally consider one non-BPS D8-brane,
which can be realized in perturbative type I string theory,
since we would like to analyze each non-BPS D8-brane
individually.
Taking T-duality transformation along the ${\bf S}^1$, we can
show that a configuration
with a non-BPS D9-brane stretched between two O$8^-$-planes
is equivalent to that with two $\wt{\mbox{O}8}^-$-planes.

\begin{figure}[htb]
\begin{center}
  \leavevmode
  \begin{picture}(250,150)
  \epsfxsize=80mm
    \put(10,0){\epsfbox{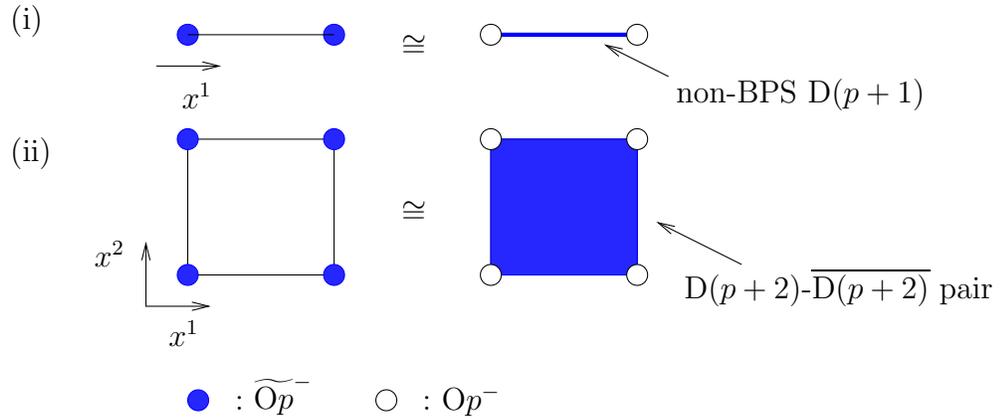}}
    \put(-40,145){(i)}
    \put(-40,95){(ii)}
    \put(108,135){$\cong$}
    \put(108,72){$\cong$}
    \put(25,115){$x^1$}
    \put(20,25){$x^1$}
    \put(-8,55){$x^2$}
    \put(45,1){:~$\wt{\mbox{O}p}^-$}
    \put(116,1){:~O$p^-$}
    \put(212,117){non-BPS D$(p+1)$}
    \put(215,43){D$(p+2)$-\AD{(p+2)} pair}
  \end{picture}
\end{center}
  \caption{\small
  The relations (i) and (ii). The types of
  O$p$-planes are flipped after condensation of the tachyonic
  modes on the `non-BPS D-branes'. }
  \label{zu:nonB}
\end{figure}

This can be generalized as follows.
Let us compactify the $x^1$, $x^2$,$\ldots$ and $x^{9-p}$ directions
of type I string theory
on ${\bf T}^{9-p}$ and fix $O(32)$ Wilson lines
to arbitrary values.
To introduce a non-BPS D8-brane, we add a half D9-brane and a half
\AD9-brane
with the following Wilson lines.
\begin{equation}
g_1=-1,\quad g_2=\cdots=g_{9-p}=+1,\quad \mbox{(for D9)}
\end{equation}
\begin{equation}
g_1'=g_2'=\cdots=g_{9-p}'=+1.\quad \mbox{(for \AD9)}
\end{equation}
If we carry out the T-duality transformation before the tachyon
condensation,
these Wilson lines change the charge of an O$p$-plane at
$(x^1,x^2,\ldots,x^{9-p})=(+1,+1,\ldots,+1)$ by $-1/2$
and one at $(-1,+1,\ldots,+1)$ by $+1/2$.
(We normalize the coordinate such that the
orientifold planes are located at $x^i=\pm1$.)
On the other hand, if we take a T-duality transformation after the
tachyon condensation,
a non-BPS D8-brane is mapped to a non-BPS D$(p+1)$-brane stretched
between
the two orientifold planes
(Fig.\ref{zu:nonB}(i)).
Therefore, we obtain the following relation.
\begin{quote}
\underline{Relation (i)}\\
\it{A configuration with a non-BPS D$(p+1)$-brane stretched between
two O$p$-planes is topologically equivalent to that
with two opposite type O$p$-planes.}
\end{quote}
Here we only consider O$p^-$ and $\wt{\mbox{O}p}^-$, and
they are called the opposite type to each other in the above statement.
This relation is essentially shown in \cite{transfer} and
used in the explanation of the brane transfer operation.
Note that the non-BPS D$(p+1)$-brane above is unstable due to
tachyonic modes on it.
In usual situations in type II string theory
without orientifold planes, non-BPS D-branes do not have
any conserved charge and they are believed to decay into the
supersymmetric vacuum after the tachyon condensation.
However, in our case,
the non-BPS D$(p+1)$-brane changes the type of orientifold plane
which the brane is attached on.
This implies that non-BPS D$(p+1)$-branes carry a ${\bf Z}_2$ charge
as well as type I D8-branes.

Next, we would like to consider T-dual picture of
type I non-BPS D7-branes.
Let us consider type I string theory compactified on ${\bf T}^{9-p}$
and fix the $O(32)$ Wilson lines to arbitrary values again.
A non BPS D7-brane perpendicular to the $x^1$-$x^2$ plane
is realized by adding two (four halves) D9-branes and two (four halves)
\AD9-branes
with the following Wilson lines.
\begin{equation}
\left.
\begin{array}{c}
g_1=\diag(-1,-1,+1,+1),\quad g_2=\diag(-1,+1,-1,+1),\\
g_3=\cdots=g_{9-p}=\diag((+1)^4).
\end{array}
\right\}\quad \mbox{(for D9)}
\label{WilB}
\end{equation}
\begin{equation}
g_1'=g_2'=\cdots=g_{9-p}'=\diag((+1)^4).\quad \mbox{(for \AD9)}
\end{equation}
The Wilson lines (\ref{WilB}) are also argued
to be unacceptable in section \ref{Wil.sec}.
Here, let us formally consider these Wilson lines
as we did in the case of the non-BPS D8-brane above.
If we carry out the T-duality transformation before taking account of
the tachyon condensation,
these extra Wilson lines change the charges of four O$p$-planes
at $(\pm1,\pm1,(+1)^{7-p})$ by $-3/2$, $+1/2$, $+1/2$, $+1/2$,
respectively.
(The integral part of this charge assignment is not important because
we can change them by moving D$p$-branes from one O$p$-plane to
another.)
Therefore, the type of these four O$p$-planes are changed.
On the other hand, if we first take the tachyon condensation on the
D9-\AD9 system
to obtain a non-BPS D7-brane perpendicular to the $x^1$-$x^2$ plane
and then carry out the T-duality transformation for all directions on
${\bf T}^{9-p}$,
we obtain one `non-BPS D$(p+2)$-brane' wrapped on a hyperplane
containing the
four O$p$-planes
at $(\pm1,\pm1,(+1)^{7-p})$.
Because a type I non-BPS D7-brane can be constructed as
a D7-\AD7 pair in type IIB string theory
projected by $\Omega$, where $\Omega$ is the world-sheet parity
operator exchanging D7 and \AD7  \cite{WitKO,FGLS},
the `non-BPS D$(p+2)$-brane' is nothing but a pair of
a D$(p+2)$ and a \AD{(p+2)}.
Hence we conclude that
(Fig.\ref{zu:nonB}(ii)):
\begin{quote}
\underline{Relation (ii)}\\
\it{A configuration with a D$(p+2)$-\AD{(p+2)} pair stretched among
four O$p$-planes is topologically equivalent to that
with four opposite type O$p$-planes.}
\end{quote}
Note again that
the D$(p+2)$-\AD{(p+2)} pair has tachyonic modes
apart from the O$p^-$-planes.
After condensation of these tachyonic modes,
the types of four O$p$-planes on the `non-BPS D$(p+2)$-brane'
are flipped.
This reflects the fact that the `non-BPS D$(p+2)$-brane'
carries a ${\bf Z}_2$ charge,
which corresponds to the non-BPS D7-brane charge in
type I string theory.

In addition to these T-duality relations among
non-BPS configurations,
they are related by `the descent relation' \cite{descent,nonbps},
which relates D-branes with different dimensions via tachyon
condensations.
On the type I side, we can construct a non-BPS D7-brane using two
non-BPS D8-branes.
Let us compactify type I string theory on ${\bf S}^1$ and
consider two non-BPS D8-branes wrapped on the ${\bf S}^1$.
Since the gauge group on a non-BPS D8-brane is
${\bf Z}_2$\cite{transfer},
we can assign different Wilson lines for two non-BPS D8-branes.
Due to these Wilson lines, tachyon condensation generates a kink
solution and it is identified with a non-BPS D7-brane.
On type II side,
we can construct non-BPS D$(p+1)$-branes stretched between two
O$p$-planes from a D$(p+2)$-\AD{(p+2)} pair.
Let us consider a  D$(p+2)$-\AD{(p+2)}
pair wrapped on a hyperplane containing four O$p$-planes
at $(\pm1,\pm1,(+1)^{7-p})$.
There are two real tachyon fields on the D$(p+2)$-\AD{(p+2)} pair.
Let $T(x^2)$ denote a mode of one of these tachyon fields
depending only on $x^2$.
Because the tachyon field changes its sign by the orientifold flip,
$T(x^2)$ cannot have an expectation value at $x^2=\pm1$.
Therefore, as a result of the condensation of the $T$-field,
we obtain a pair of kink solutions at $x^2=\pm1$.
These are identified with two non-BPS D$(p+1)$-branes, which are
T-dual to the two non-BPS D8-branes
used to construct a non-BPS D7-brane.
Similarly, we can proceed one more step.
There is a real tachyon field on the non-BPS D$(p+1)$-brane,
whose sign is flipped under the orientifold flip.
Hence, we obtain a pair of kink solutions at the locations of
the two O$p$-planes, $x^1=\pm1$, as above.
These are now identified with two 1/2 D$p$-branes stuck on
the two O$p$-planes.

Now let us consider type I string theory compactified on ${\bf T}^3$
which is parameterized by $x^1$,$x^2$ and $x^3$, and investigate
more detailed structure. We study the dual ${\bf T}^3/{\bf Z}_2$
orientifold for simplicity,
but most of the results below can easily be generalized to
${\bf T}^n/{\bf Z}_2$ orientifolds.

\begin{figure}[htb]
  \leavevmode
  \epsfxsize=80mm
  \begin{center}
  \begin{picture}(250,150)
  \put(10,0){\epsfbox{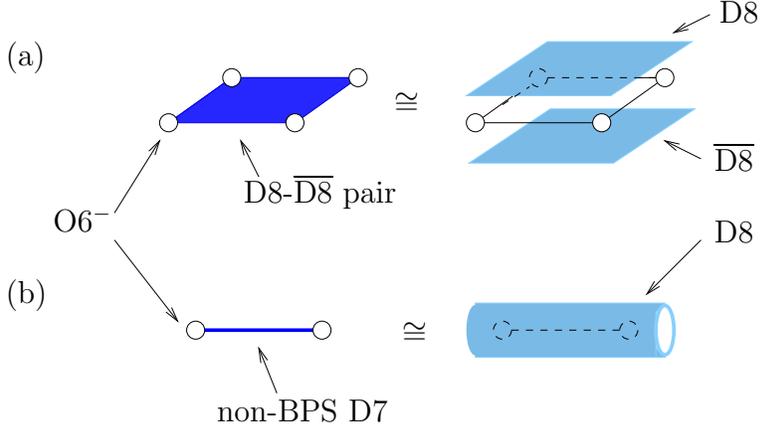}}
    \put(-30,125){(a)}
    \put(-30,35){(b)}
    \put(117,107){$\cong$}
    \put(120,20){$\cong$}
    \put(238,85){\AD8}
    \put(240,141){D8}
    \put(238,58){D8}
    \put(60,73){D8-\AD8 pair}
    \put(50,-10){non-BPS D7}
    \put(-12,62){O$6^-$}
  \end{picture}
  \end{center}  
  \caption{\small
  Descriptions of the `non-BPS D-branes'
  in ${\bf T}^3/{\bf Z}_2$ orientifold using D8-branes.
  The outside region
  divided by the D8-branes is in the background
  with odd cosmological constant.
  }
  \label{zu:nonD8}
\end{figure}

As shown in \cite{WitKO,transfer}, 
the gauge group of a type I non-BPS D7-brane is $U(1)$.
Let us consider a non-BPS D7-brane which is perpendicular to 
the $x^1$ and $x^2$ direction.
Taking T-duality along the $x^1$, $x^2$ and $x^3$ directions,
we obtain a D8-\AD8 pair which is localized in the $x^3$
direction and extended in the $x^1$ and $x^2$ directions.
The degree of freedom of a $U(1)$ Wilson line along $x^3$
on the non-BPS D7-brane
implies that the D8-\AD8 pair in the T-dualized picture
can split and move in the $x^3$ direction.
Therefore, we obtain a picture that
four O$6^-$-planes are caught between a
 D8-brane and a \AD8-brane in the covering space.
(Fig.\ref{zu:nonD8}(a))

Similarly, a non-BPS D7-brane stretched between two O$6^-$-planes
can be interpreted as
a configuration with two O$6^-$-planes cylindrically wrapped
by a D8-brane. (Fig.\ref{zu:nonD8}(b))
In fact,  the non-BPS D7-brane can be realized as a kink configuration
of a D8-\AD8 pair, in which the tachyons created by
the D8-\AD8 strings are condensed apart from a hyperplane of
codimension one stretched between two O$6^-$-planes.
If we identify the tachyon condensation with the pair annihilation of
the D8-\AD8 pair,
we actually obtain the cylindrical configuration.

Using this re-interpretation of non-BPS D-branes by D8-branes,
we can easily explain why these non-BPS D-branes
change the type of O$6$-planes when they decay.
The D8-branes in Fig.\ref{zu:nonD8} can be deformed into
spherical D8-branes enclosing O$6$-planes.
According to the argument in section \ref{disc.sec},
these spherical D8-branes change the type of O$6$-planes.

Furthermore, we can argue the shifted quantization condition
(\ref{quant}) in terms of the tachyon condensation.
Consider an O$6$-plane located at $x^1=x^2=x^3=0$
and a D$8$-\AD{8} pair  on top of it, localized in the $x^3$ direction.
There is a complex tachyon field denoted again by $T$
on the world-volume of the D$8$-\AD{8} pair.
We consider the tachyon field which depends only on $x^1$ and $x^2$.
As noted above, this tachyon field changes its sign by the
orientifold flip and hence satisfies
\begin{eqnarray}
T(-x^1,-x^2)=-T(x^1,x^2)
\label{tachyon}
\end{eqnarray}
in the covering space.
Note that this equation implies $T(0,0)=0$.
Suppose that the tachyon condenses apart from the O6-plane, namely,
$T(x^1,x^2)\ne 0$ for all $(x^1,x^2)\ne(0,0)$.
Then it is easy to show using (\ref{tachyon}) that it makes a vortex
of odd winding number. This implies that it behaves as an odd number
of half D$6$-branes. If we think that the D$8$-\AD{8} pair is
annihilated apart from the O$6$-plane due to the tachyon condensation,
we will have a configuration with the O$6$-plane wrapped by a
spherical D$8$-brane. Therefore, what we have shown here
is nothing but the shifted quantization condition (\ref{quant}).

Furthermore, we can easily select allowed 
configurations including non-BPS D7-branes and/or D8-\AD8 pairs
with a help of the re-interpretation.
A D8-brane
changes the background cosmological constant by one unit.
Therefore, if we accept the fact suggested in section \ref{tdual.sec}
that O$6^-$($\wt{\mbox{O}6}^-$)-planes are allowed only in the
background with even (odd) cosmological constant,
we can easily show that the configurations associated with
the Wilson lines (\ref{WilA})
and (\ref{WilB}) are not allowed.
The allowed configurations
can always be transformed to the configuration
with eight O$6^-$-planes or that with eight $\wt{\mbox{O}6}^-$-planes
using relations (i) and (ii).

\begin{figure}[htb]
  \leavevmode
  \epsfxsize=130mm
  \begin{center}
  \begin{picture}(600,70)
  \put(25,0){\epsfbox{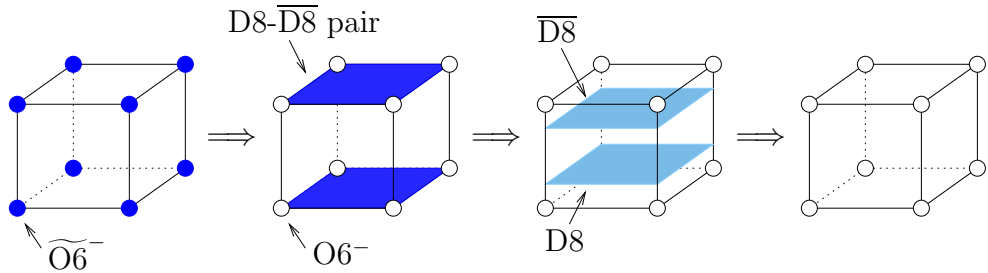}}
    \put(225,79){\AD8}
    \put(228,0){D8}
    \put(108,83){D8-\AD8 pair}
    \put(40,-7){$\wt{\mbox{O}6}^-$}
    \put(140,-5){O$6^-$}
    \put(100,40){$\Longrightarrow$}
    \put(200,40){$\Longrightarrow$}
    \put(300,40){$\Longrightarrow$}
  \end{picture}
  \end{center}
  \caption{\small
  Transfer of stuck D-branes.
  The system with eight $\wt{\mbox{O}6}^-$-planes can be
  continuously deformed to that with eight O$6^-$-planes.}
  \label{zu:trans}
\end{figure}

Finally, we would like to point out another phenomenon
concerned with the transfer of stuck D-branes.
Let us consider type I string theory compactified on ${\bf T}^3$
with Wilson lines (\ref{T3WL}), and take
T-duality along all three directions of the torus.
Namely we consider ${\bf T}^3/{\bf Z}_2$ orientifold with
eight $\wt{\mbox{O}6}^-$-planes.
{}Using the relation (ii), we know that
this system is topologically equivalent to
that with eight O$6^-$-planes and two D8-\AD8 pairs, as depicted
in Fig.\ref{zu:nonB}(ii).
Now, recall that the D8-\AD8 pair can
split and move along the $x^3$ direction. 
If we move one of the D8-\AD8 pairs to coincide with the other,
they will annihilate completely and only eight O$6^-$-planes
remain. (Fig.\ref{zu:trans})
Namely, the system with eight $\wt{\mbox{O}6}^-$-branes
can be continuously deformed to that
with eight O$6^-$-branes!
We will examine this process more carefully
in the next section.

%%%%%%%%%%%%%%%%%%%%%%%%%%%%%%%%%%%%%%%%%%%%%%%%%%%%%%%%%%%%%%%%%%%%%%
\section{Transfer of Stuck D-branes}\label{transfer.sec}
In section \ref{disc.sec} we showed that
stuck D$p$-branes on O$p$-planes
are not really stuck but can be moved away
from O$p$-planes as magnetic flux on D$(p+2)$-branes.
Using this, we can continuously deform an orientifold
with eight O$6^-$-planes to that with eight $\wt{\mbox{O}6}^-$-planes
as mentioned in the previous section.
In the context of ${\bf T}^3$ compactified type I string theory,
this implies the trivial Wilson lines can be continuously deformed
into the Wilson lines (\ref{T3WL}).
The purpose of this section is to show that this is actually possible
and that the process is actually related to a deformation of D8-branes
in ${\bf T}^3/{\bf Z}_2$ orientifold via T-duality.

Let us consider Wilson lines of gauge group $G=Spin(8)$ on
${\bf T}^3$.
We will use only $Spin(8)$ subgroup of $Spin(32)/{\bf Z}_2$ associated
with four D-branes (and their mirror images),
and other D-branes will be omitted.
Let us start from trivial Wilson lines $g_1=g_2=g_3=1$.
We can continuously deform them
to the following Wilson lines keeping it a zero-energy configuration.
\begin{eqnarray}
g_1&=&\diag(+1,-1,+1,-1,+1,-1,+1,-1),\nonumber\\
g_2&=&\diag(+1,+1,-1,-1,+1,+1,-1,-1),\nonumber\\
g_3&=&\diag(+1,+1,+1,+1,+1,+1,+1,+1).
\label{WL1}
\end{eqnarray}
The final configuration we want to realize by continuous deformation
is
\begin{eqnarray}
g_1&=&\diag(+1,-1,+1,-1,+1,-1,+1,-1),\nonumber\\
g_2&=&\diag(+1,+1,-1,-1,+1,+1,-1,-1),\nonumber\\
g_3&=&\diag(+1,+1,+1,+1,-1,-1,-1,-1).
\label{WL2}
\end{eqnarray}
To get the final configuration (\ref{WL2})
we have to change $g_3$ in (\ref{WL1}) to that in (\ref{WL2}).
We cannot do this without raising the energy.
Let us parameterize $g_3$ between (\ref{WL1}) and (\ref{WL2}) by
$0\leq\alpha\leq1$.
The final value $g_3(\alpha=1)$ breaks the gauge group $G$ into
\begin{equation}
H
=Spin(4)^2/{\bf Z}_2
=(Spin(4)\times SU(2)\times SU(2))/{\bf Z}_2.
\end{equation}
In this subgroup, $g_3(1)$ is represented as
$g_3(1)={\bf1}_4\otimes{\bf1}_2\otimes-{\bf1}_2$.
Let us take the following path between $\alpha=0$ and $1$.
\begin{equation}
g_3(\alpha)={\bf1}_4\otimes{\bf1}_2\otimes e^{\pi i\alpha\sigma_z}.
\end{equation}
At an interpolating value of $\alpha$, the gauge group is
further broken into
\begin{equation}
H'=(Spin(4)\times SU(2)\times U(1))/{\bf Z}_2.
\label{hprime}
\end{equation}
This subgroup does not contain $g_1$ and $g_2$ as its elements.
So, when we turn on the parameter $\alpha$,
the gauge configuration on the $x^1$-$x^2$ plane must be deformed
such that it is embedded in the subgroup $H'$.

We can determine the gauge configuration realized after turning on
$\alpha$ as follows.
Let us regard $g_1$ and $g_2$ as elements of $H$.
The non-trivial homotopy $\pi_1(H)={\bf Z}_2$ admits Dirac strings,
and $g_1$ and $g_2$ represent a gauge configuration with one Dirac
string.
Even if the gauge group is broken to $H'$,
the Dirac string should remain.
More precisely, $\pi_1(H')={\bf Z}$ due to the $U(1)$ factor
and the number of Dirac strings of gauge group $H'$ should be an odd
integer.
Unlike $SU(2)$ case, gauge configuration with $U(1)$ Dirac strings
cannot be realized as a vacuum configuration.
The Dirac strings must always be accompanied by the same number of
magnetic flux.
(For example, the magnetic charge of a Dirac monopole is equal to the
number of
Dirac strings going out from the monopole.)
Therefore, when $0<\alpha<1$,
nonzero magnetic flux associated with the $U(1)$ factor of $H'$ is
generated.

When $\alpha$ reach the final value $1$, the gauge group $H$ is
restored and
$U(1)$ factor is enhanced to $SU(2)$ again.
At the same time, the magnetic flux vanishes to leave the Wilson lines
$g_1$ and $g_2$ again
and finally we get the Wilson lines (\ref{WL2}).

Next, let us discuss T-dual of this process.
Before going to ${\bf T}^3/{\bf Z}_2$ orientifold,
it is convenient to insert one step.
By taking T-duality only along the $x^3$-direction for type I string
theory,
we have ${\bf S}^1/{\bf Z}_2\times{\bf T}^2$ orientifold
in type IIA string theory.
This contains four D8-branes (and $12$ we are now omitting).
At first, when $\alpha=0$, these D8-branes stay on one
of two O8-planes.
The gauge group $G=Spin(8)$ is realized on this O8-plane.
Turning on the parameter $\alpha$ corresponds
to moving two of these D8-branes from the O8-plane.
Each factor of the broken gauge group $H'\sim Spin(4)\times U(2)$
represent
the gauge symmetries on two D8-branes staying on the O8-plane
and two moving D8-branes respectively.

When $\alpha\neq0$,
the non-zero magnetic flux in type I configuration is mapped into
magnetic flux on the moving D8-branes.
The magnetic flux on the D8-branes can be regarded as D6-branes
absorbed in the world-volume of the D8-branes, and
by checking the amount of the flux carefully
we can find the number of the D6-branes is odd integer.

By taking further T-duality along the $x^1$ and $x^2$ directions
for this configuration,
we have a dual configuration with
odd number of D8-branes wrapped along $x^1$ and $x^2$ directions
and they hold two D6-branes on them.
This actually shows that stuck D6-branes are transferred
as magnetic flux on the D8-branes.
The emergence and annihilation of the $U(1)$ magnetic flux
correspond to the pair creation and pair annihilation of
D8-branes and \AD8-branes, which are mirror images of the D8-branes.
This is the same relation with what used in (\cite{AHH}) to analyze
brane annihilation in terms of matrix theory.

%%%%%%%%%%%%%%%%%%%%%%%%%%%%%%%%%%%%%%%%%%%%%%%%%%%%%%%%%%%%%%
\section{Conclusion and Discussion}
In section \ref{Wil.sec} and \ref{tdual.sec}
we focused on Wilson lines in type I string theory
with vector structure
and showed that the constraints for the Wilson lines are
closely related to the discrete torsion of O$p$-planes associated
with  the R-R $(5-p)$ form field.
It would be an interesting problem to extend this analysis to Wilson
lines without vector structure.
It was partially done in \cite{keur} up to ${\bf T}^3$
compactification.
Because it is known that such Wilson lines
are related with O$p^+$ and $\wt{\mbox{O}p}^+$-planes\cite{novector},
if we can obtain such a set of constraints for general Wilson lines,
we would obtain more information about properties of O$p^+$ and
$\wt{\mbox{O}p}^+$.
Furthermore, it is also interesting
to consider a relation between type I Wilson lines and
extra non-trivial cohomologies for lower dimensional orientifold
planes.

In section \ref{disc.sec} we showed wrapped branes change the discrete
torsions of orientifold planes.
To explain the change of the R-R charges of the orientifold planes,
we used the shifted flux quantization conditions on the branes
wrapped around the orientifold planes.
As explained in section \ref{non-bps.sec},
we obtained an interesting derivation of the condition (\ref{quant}),
but this is not the end of the story.
We do not have any proof of the condition (\ref{hquant})
as well as the explanation for the fractional charges of orientifold
planes, which may be closely related to the
quantization condition on the wrapped branes.
Instead, we just notice that
we cannot naively conclude they are inconsistent with the Dirac's
quantization condition.
If one want to obtain the quantization condition,
one should use a brane coupling to the gauge flux.
Deforming the brane along a closed path in the configuration space and
demanding two partition functions for initial and final
brane configurations to coincide,
the Dirac's condition for the background gauge flux is
obtained.
In string theory, however, this is not a simple task because
such branes are accompanied by dynamical fields on it.
To obtain the correct partition functions, we should trace how
these fields on branes vary during the brane deformation.
Recently, such analysis was done for some cases
in several works \cite{anomalies,flux} and similar approach is
expected
to solve the problem about the fractional charges of orientifold
planes and shifted quantization conditions.

In this paper, we have not paid much attention to the supersymmetry.
Because our analysis relies on topological aspects,
we expect our arguments are reliable in non-supersymmetric cases.
However, it is worth asking whether each Wilson line
can be realized as a supersymmetric configuration.
Concerning this question, there is one subtlety.
In section \ref{Wil.sec} we required the commutativity of type I
Wilson lines.
In ordinary Yang-Mills theory, this requirement guarantees
the vanishing field strength and the vanishing energy.
In supergravity, however, this is not sufficient to obtain a
zero-energy configuration
because the field strength of the two-form field in type I string
theory contains the Chern-Simons invariant.
\begin{equation}
H_3=dB_2-\omega_3.
\end{equation}
Even in the cases of vanishing Yang-Mills field strength,
the Chern-Simons invariant
does not always vanish.
If it takes some fractional value, it cannot be canceled by the $dB_2$
term.%
\footnote{Y. I. thanks J. deBore and K. Hori
for pointing out this fact.}
Actually, the Wilson lines (\ref{T3WL}) gives $\int\omega_3\in
{\bf Z}+1/2$
(The integral part is not determined even if we specify the
Wilson lines because
$\int\omega_3$ is not invariant under large gauge transformations.)
and we cannot cancel this by $\int dB_2$, which is always integer.
This corresponds to the fact that the configuration with
eight $\wt{\mbox{O}6}^-$-planes has odd cosmological constant.
Via T-duality, $H_3$ is related with the cosmological constant
$\Lambda$
on the ${\bf T}^3/{\bf Z}_2$ orientifold by the equation,
\begin{equation}
\int H_3=\frac{1}{2}\Lambda.
\end{equation}
Notice that the normalization of the $H_3$-flux differs from that of
R-R fluxes
in type II string theory by factor $2$ because type I D1-branes are
fractional branes stuck on an O$9^-$-plane.
Therefore, we cannot realize the Wilson lines (\ref{T3WL}) on a flat
spacetime.
This, however, does not implies inconsistency of the theory unlike
anomalies.
It just means the lower dimensional theory obtained by the
compactification
has nonzero cosmological constant.
Indeed, we can make a static and even supersymmetric classical
solution like the type I' configuration.

\section*{Acknowledgments}
Y.H. and S.S. would like to thank their colleagues at YITP for valuable
discussions and encouragement.
Y.I. is grateful to A.Hanany for discussions.
The work of S.S. is supported in part
by the Grant-in-Aid for JSPS fellows.
The work of Y.I. was supported in part by funds provided by
the U.S. Department of Energy (D.O.E.) under cooperative research agreement
\#DE-FC02-94ER40818 and by a Grant-in-Aid for Scientific Research from
the Ministry of Education, Science, Sports and Culture (\#9110).
%%%%%%%%%%%%%%%%%%%%%%%%%%%%%%%%%%%%%%%%%%%%%%%%%%%%%%%%%%%%%%

\end{document}